\documentclass[12pt]{article}
\usepackage{natbib}
\usepackage{amsmath,amssymb,mathtools}
\usepackage{newtxtext,newtxmath}
\usepackage{geometry}
\usepackage{graphicx}
\usepackage{float}
\usepackage{algorithm}
\usepackage{algpseudocode}
\usepackage{booktabs}
\usepackage{hyperref}
\usepackage{multirow}
\usepackage{setspace}
\usepackage{authblk}
\usepackage{subcaption}
\usepackage{xcolor}

\title{Machine Learning-Enabled Time-Skipping Method: Theory and
Perturbation-Based Verification}

\author[1]{Isaac C. Bannerman}
\author[1]{Shaowu Pan}
\author[1]{Lucy T. Zhang\thanks{Corresponding author: zhanglucy@rpi.edu}}
\affil[1]{Department of Mechanical, Aerospace, and Nuclear Engineering,
Rensselaer Polytechnic Institute, 110 8th Street, Troy, NY 12180, USA}

\date{}

\begin{document}

\maketitle

\begin{abstract}

Simulating the time evolution of practical mechanics problems requires many time steps. This cost is further compounded by the iterative nature of implicit solvers. We present a cycle-based \textbf{M}achine \textbf{L}earning-enabled \textbf{T}ime-skipping (MLT) method. In each cycle, several time steps are skipped, a machine learning (ML) prediction bridges the gap to initialize one solver step, and the physics-based solver then runs for several steps to correct any accumulated errors. Using the Implicit Function Theorem, we derive first-order estimates that characterize how ML prediction errors propagate. These estimates identify when errors remain bounded across cycles and guide the choice of cycle
parameters, independent of model architecture. We verify these estimates on the Allen--Cahn equation and a forced variant using controlled synthetic perturbations that emulate ML errors. Wall-clock measurements confirm the theoretical speedup is realizable in practice. 
The MLT method provides a robust framework for accelerating time-marching in dynamic simulations. 

\end{abstract}

\vspace{0.5em}
\vspace{1em}

\newpage
\section{Introduction}
Simulating the time evolution of practical mechanics problems --- including fluid flows~\cite{moin1998direct, yu2022stabilized}, weather prediction~\cite{lynch2008origins, bauer2015quiet, bi2023accurate}, and fluid-structure interaction (FSI)~\cite{liu2006immersed, hou2012numerical, 
sun2025parallel} --- requires a large number of iterations to accurately resolve the underlying physics. Explicit time integrators are constrained by the Courant--Friedrichs--Lewy (CFL) condition~\cite{cfl}, requiring impractically small time steps. Implicit methods overcome this, but introduce nested iterations at every step: Newton linearization, Krylov solves~\cite{knoll2004jacobian, vorst2003iterative, saad1986gmres}, and preconditioner sub-iterations~\cite{saad1993flexible}. Since time stepping sits at the top of this hierarchy, reducing the number of time steps yields a cascading reduction in computational work across all levels. The computational burden of these nested iterations has motivated a growing body of research on data-driven acceleration methods. Rather than replacing physics-based solvers entirely, these methods combine fast ML-based approximations with them to achieve both speed and 
reliability~\cite{wang2024artificial, kochkov2021machine}. We organize existing work by level of the nested hierarchy.

At the linear solver level, several works have explored using learned models to accelerate Krylov iterations. Hsieh et al.~\cite{hsieh2019learning} proposed modifying solver update steps via a neural network while preserving convergence guarantees. Bin et al.~\cite{bin2025data} demonstrated significant speedups by projecting solutions onto a low-dimensional subspace. Bai and Bian~\cite{bai2026hybrid} used a DeepONet~\cite{lu2021learning} to predict search directions for conjugate gradient solvers in incompressible flows.
At the Newton iteration level, the focus shifts to reducing Newton steps through better initializations or preconditioners. Aghili et al.~\cite{steady_newton_accel} used Fourier Neural 
Operators~\cite{Fourier} to accelerate Newton iterations for nonlinear elliptic problems. Lee et al.~\cite{lee2025neural} proposed a neural-operator preconditioned Newton method, demonstrating robustness across problems with strong nonlinearities. Jin et al.~\cite{ml_for_allen_cahn} reduced Newton iterations in stiff parabolic problems using improved initializations from a neural network.

At the time-stepping level --- the highest and most impactful level --- fewer works have directly targeted a reduction in the number of time steps. For steady-state problems, data-driven warm-starting approaches have been used to reduce the number of pseudo time steps needed to reach convergence. 
Obiols-Sales et al.~\cite{obiols2020cfdnet} and Zhang et al.~\cite{CFDNet_using_graph} used neural networks to generate approximate flow fields that initialize the solver. Zhou et al.~\cite{zhou2025neural} introduced a super-fidelity warm-start approach, mapping low-fidelity 
solutions to high-fidelity ones, accelerating steady-state solvers. Wang et al.~\cite{wang2026data} periodically solved a least-squares problem in a reduced-order subspace to accelerate convergence. For transient problems, Sharpe et al.~\cite{sharpe2025accelerating} reduced the number of time 
steps by replacing traditional initial conditions with inexpensive ML-predicted flow fields. Eshaghi et al.~\cite{eshaghi2025nows} accelerated Krylov solvers at each time step through learned Krylov initializations. Despite this progress, directly skipping time steps in dynamic simulations remains largely unexplored. Skipping steps introduces errors that propagate and accumulate across the simulation horizon, making it more challenging than warm-starting. Yet each skipped step eliminates an entire 
Newton--Krylov solve, making time-step skipping the highest-leverage target for acceleration.

We propose MLT, a cycle-based machine-learning-enabled time-skipping method. In a given cycle, some time steps are skipped, and an ML model predicts the system state after the skip. The physics-based solver then takes this prediction as a starting point for one step, rather than the true state. The solver then continues without ML input for several more steps, correcting any error the prediction introduced. 
We analyze this design using the Implicit Function Theorem~\cite{krantz2002implicit, rudin1976principles,
dontchev2009implicit}, deriving first-order estimates of how prediction errors propagate. These estimates identify when errors remain bounded across cycles; they also guide the choice of skip length and correction window. Because they are independent of model architecture or training procedure, the analysis specifies the accuracy an ML predictor must achieve before one is trained. We verify these estimates on the Allen--Cahn equation and a forced variant using controlled synthetic perturbations. Wall-clock measurements confirm that the theoretical speedup is realizable in practice.

The remainder of this paper is organized as follows. Section~\ref{sec:method} describes the MLT method. Section~\ref{sec:theory} develops the error propagation theory. 
Section~\ref{sec:verification} verifies the theoretical predictions on the Allen--Cahn equation and quantifies the wall-clock speedup. Section~\ref{sec:conclusion} concludes and gives future directions.


\section{The Methodology}\label{sec:method}

\subsection{A Quick Review of Time Integration Schemes}\label{sec:timestep}

Practical mechanics problems are governed by nonlinear dynamical 
systems which, after spatial discretization, take the semi-discrete form
\begin{equation}
    \frac{d\mathbf{x}}{dt} = \mathbf{G}(\mathbf{x}),
    \label{eq:semidiscrete}
\end{equation}
where $\mathbf{x} \in \mathbb{R}^N$ is the state vector and
$\mathbf{G} : \mathbb{R}^N \to \mathbb{R}^N$ is a nonlinear
operator arising from the spatial discretization of the underlying partial differential equation (PDE).
Applying a time discretization with step size $\tau$ to
Eq.~\eqref{eq:semidiscrete} yields a discrete update relation between
consecutive states $\mathbf{x}^{t-1}$ and $\mathbf{x}^t$. The precise
form of this relation depends on the chosen time integration scheme. Explicit schemes compute the new state directly:
\begin{equation}
    \mathbf{x}^{t} = \mathbf{x}^{t-1} + \tau\,
    \mathbf{G}(\mathbf{x}^{t-1}),
    \label{eq:explicit}
\end{equation}
as in forward Euler or Runge--Kutta schemes~\cite{hairer1993solving}, 
but are subject to the CFL stability condition~\cite{cfl}. However, implicit schemes define the new state through a nonlinear residual:
\begin{equation}
    F(\mathbf{x}^{t-1},\, \mathbf{x}^{t}) = \mathbf{0},
    \label{eq:implicit_system}
\end{equation}
where $F$ encodes the time-differencing stencil and the implicit 
spatial operator. For the implicit midpoint method:
\begin{equation}
    F(\mathbf{x}^{t-1}, \mathbf{x}^{t})
    = \mathbf{x}^{t} - \mathbf{x}^{t-1}
      - \tau\,\mathbf{G}\!\left(\frac{\mathbf{x}^{t-1}+
      \mathbf{x}^{t}}{2}\right).
    \label{eq:midpoint_residual}
\end{equation}

Implicit schemes overcome the CFL restriction but require solving a nonlinear system Eq.~\eqref{eq:implicit_system} at every step~\cite{hairer1996solving} typically via a Newton--Krylov method~\cite{knoll2004jacobian}. Given the current iteration solution $\mathbf{x}_k^t$, Newton's method linearizes
the residual and solves
\begin{equation}
    \nabla_{\mathbf{x}^t} F(\mathbf{x}^{t-1},\, \mathbf{x}_k^t)\,
    \delta\mathbf{x}_k^t
    = -F(\mathbf{x}^{t-1},\, \mathbf{x}_k^t),
    \label{eq:newton_step}
\end{equation}
for the update $\delta\mathbf{x}_k^t$, after which the solution is
advanced as $\mathbf{x}_{k+1}^t = \mathbf{x}_k^t +
\delta\mathbf{x}_k^t$. The Jacobian-vector products in Eq.~\eqref{eq:newton_step} are typically computed matrix-free using a Krylov solver such as the Generalized Minimal 
Residual method (GMRES)~\cite{saad2003iterative}, yielding a 
Jacobian-free Newton--Krylov framework~\cite{knoll2004jacobian}. Starting from $\mathbf{x}_0^t = \mathbf{x}^{t-1}$, the Newton 
iteration is repeated until $\|F(\mathbf{x}^{t-1}, 
\mathbf{x}_k^t)\| < \epsilon$ for a prescribed tolerance $\epsilon$, 
giving the converged solution $\mathbf{x}^t$. In 
many practical problems, a preconditioner $\mathbf{P}^{-1} \approx 
[\nabla_{\mathbf{x}^t} F]^{-1}$ is applied to accelerate Krylov 
convergence. For preconditioners such as those used in multigrid or 
FSI methods~\cite{washio2005robust}, $\mathbf{P}^{-1}$ itself requires an iterative sub-solve. This introduces a fourth nested level of iterations beyond time stepping, Newton iterations, and Krylov iterations. The complete procedure for 
the case of an implicit solver is summarized in 
Algorithm~\ref{alg:baseline}.

Regardless of the scheme, each time step can be viewed as a single application of a time-step operator. This abstraction allows the MLT framework to be stated in a scheme-agnostic way.

\begin{algorithm}
\caption{Physics-based Solver (Implicit Case)}
\label{alg:baseline}
\begin{algorithmic}[1]
\State Set tolerance $\epsilon$, maximum Newton iterations $k_{\max}$, 
       maximum Krylov iterations $l_{\max}$
\For{time step $t = 1, \ldots, N$}
    \State Initialize $\mathbf{x}^t_{k=0} \leftarrow \mathbf{x}^{t-1}$
    \For{$k = 0, 1, \ldots, k_{\max}$} \Comment{Newton iterations}
        \State Evaluate residual 
               $\mathbf{r} = F(\mathbf{x}^{t-1}, \mathbf{x}^t_k)$
        \If{$\|\mathbf{r}\| < \epsilon$} \textbf{break}
        \EndIf
        \For{$l = 0, 1, \ldots, l_{\max}$} \Comment{Krylov iterations}
            \State Apply preconditioner $\mathbf{P}^{-1}$ 
                   (may itself require inner iterations)
            \State Update Krylov subspace approximation 
                   to $\delta\mathbf{x}^t_k$
            \If{Krylov residual $< \epsilon_{\text{Krylov}}$} 
               \textbf{break}
            \EndIf
        \EndFor
        \State Update: 
               $\mathbf{x}^t_{k+1} = \mathbf{x}^t_k + \delta\mathbf{x}^t_k$
    \EndFor
    \State Set $\mathbf{x}^t \leftarrow \mathbf{x}^t_k$
\EndFor
\end{algorithmic}
\end{algorithm}

\subsection{The Machine Learning-Enabled Time-Skipping (MLT) 
Method}\label{sec:mlt}

Having established the time-stepping method for nonlinear dynamical systems, we now describe the MLT method. The MLT method reduces cost by selectively skipping time steps and using a machine learning model to bridge the resulting gaps while a physics-based solver maintains physical consistency. It operates in two phases: an \textbf{initial 
warm-up stage} followed by a \textbf{cycle-based stage}.

In the warm-up stage, the physics-based solver runs for $W$ steps without ML intervention to establish a physically consistent trajectory. Once complete, the cycle-based stage begins where the simulation interval is partitioned into cycles of length $c$, each consisting of three phases:
\begin{itemize}
    \item \textbf{Skipped steps ($s$):} $s$ time steps are skipped. No 
    solver calls are made and the time index is advanced by $s$.
    \item \textbf{ML-initialized step (1):} an ML prediction initializes 
    one physics-based solver step, bridging the skipped steps and the 
    physics-based correction steps that follow.
    \item \textbf{Physics-based correction steps ($m$):} the physics-based solver then runs for $m$ consecutive time steps without any ML intervention, 
    maintaining physical consistency.
\end{itemize}
\noindent Thus, the total length of a cycle is
\begin{equation}
    c = s + 1 + m,
    \label{eq:cycle_length}
\end{equation}
and the number of cycles is $N_c = \lfloor N/c \rfloor$, where $N$ is the total number of time steps in the cycle-based stage. The cycle parameters $m$ and $s$ need not be uniform across cycles; the present work studies the fixed-cycle case for clarity of analysis.

\begin{figure}[H]
    \centering
    \hspace*{-0.85cm}
    \includegraphics[width=1.1\textwidth]{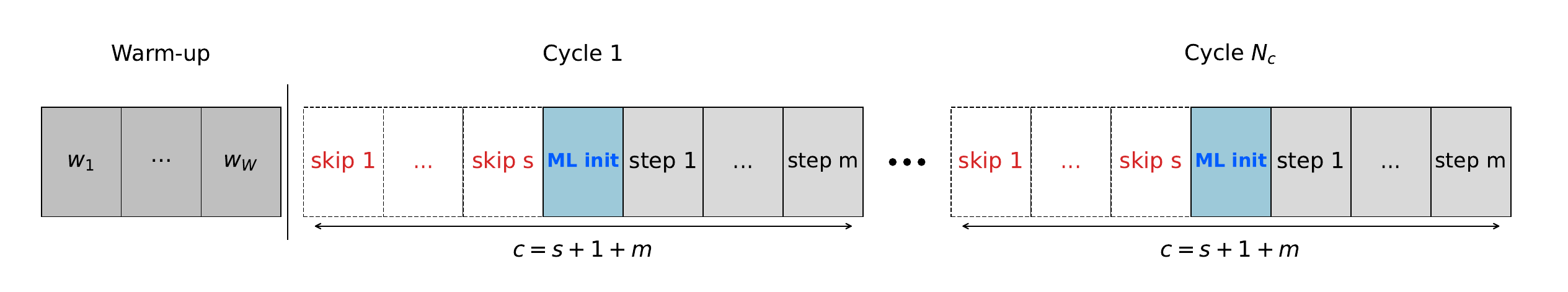}
    \caption{Schematic of multiple cycles in the MLT method. After a
    warm-up phase ($w_1, \ldots, w_W$), the physics-based solver repeats 
    cycles of length $c = s + 1 + m$, each containing $s$ skipped 
    steps, one ML-initialized step, and $m$ physics-based correction 
    steps.}
    \label{fig:multi_cycle}
\end{figure}

Introducing ML predictions into the time-stepping loop creates the risk of error accumulation across cycles. The MLT method controls this risk through a physics-based correction window and sparse ML usage; invoking the ML model only once per cycle, followed by several physics-based steps, limits how often errors are introduced and gives 
the solver time to correct them before the next cycle. The balance between $m$ and $s$ is therefore a trade-off between accuracy and savings. Larger $s$ increases acceleration but introduces larger errors, while larger $m$ improves accuracy but reduces the overall speedup. The theoretical analysis in Section~\ref{sec:theory} provides guidance for selecting these parameters. The complete MLT method is summarized in Algorithm~\ref{alg:mlt} and illustrated in Figure~\ref{fig:multi_cycle}.

\begin{algorithm}
\caption{Machine Learning-Enabled Time-Skipping (MLT) method}
\label{alg:mlt}
\begin{algorithmic}[1]
\State Set warm-up steps $W$, cycle parameters $m$, $s$
\For{the first $W$ warm-up steps}
    \State Run physics-based solver: 
           $\mathbf{x}^t \leftarrow \mathbf{x}^{t-1}$
\EndFor
\For{each cycle of length $c = s + 1 + m$}
    \State Advance time index by $s$ steps (no solver calls)
    \State Compute ML prediction: $\tilde{\mathbf{x}}^{t-1} = 
           \mathcal{NN}(\theta; \mathbf{x})$
    \State Initialize physics-based solver from $\tilde{\mathbf{x}}^{t-1}$ 
           and advance one step: $\tilde{\mathbf{x}}^t \leftarrow 
           \tilde{\mathbf{x}}^{t-1}$
    \For{$m$ physics-based correction steps}
        \State Run physics-based solver: 
               $\mathbf{x}^t \leftarrow \mathbf{x}^{t-1}$
    \EndFor
\EndFor
\end{algorithmic}
\end{algorithm}

In the numerical experiments of Section~\ref{sec:verification}, the
prediction $\tilde{\mathbf{x}}^{t-1} = \mathcal{NN}(\theta; \mathbf{x})$ is replaced by a controlled synthetic perturbation $\tilde{\mathbf{x}}^{t-1} = \mathbf{x}^{t-1} + \eta$ of prescribed magnitude. This emulates ML errors across a range of magnitudes, independent of the particular ML model or data-driven method used.

\section{Theoretical Analysis}
\label{sec:theory}
In each cycle, an ML prediction enters MLT only as a perturbation to the input state of a single solver step; the time-step operator remains the same. At every computed step, the MLT trajectory therefore coincides with the physics-based solver
when the ML prediction error vanishes. The theoretical task is to characterize how a nonzero perturbation propagates. To this end, we develop an error analysis framework based on the Implicit Function Theorem (IFT). The IFT guarantees that, under suitable conditions, the equation $F(\mathbf{a}, \mathbf{x}) = \mathbf{0}$ implicitly defines $\mathbf{x}$ as a function of $\mathbf{a}$, written $\mathbf{x} = \Phi(\mathbf{a})$. This function is locally unique and differentiable near any point where the equation holds and the Jacobian $\partial_{\mathbf{x}} F$ is nonsingular. Formulating time stepping as a nonlinear map,
we first define the error induced by a perturbation
in a single step (Section~\ref{sec:single_step}).
Building on this result, the analysis then proceeds in two stages (Section~\ref{sec:multi_step}):
first, the accumulated error over a single MLT cycle,
    obtained by applying the single-step error formula recursively
    to the ML-initialized step and the $m$ correction steps that
    follow it.
Then, the total accumulated error over the full simulation
    horizon, where new perturbations are injected at every cycle.

\subsection{Error in a Single Time Step}
\label{sec:single_step}
We begin by asking how a perturbation in the input state
of a single physics-based solver time step
contributes to the deviation in its output state.
Adopting a general nonlinear evolution problem
discretized in time using a fully implicit scheme,
two consecutive states are coupled through
a nonlinear residual relation
\begin{equation}
F(\mathbf{x}^{t-1}, \mathbf{x}^t) = \mathbf{0},
\label{eq:residual_general}
\end{equation}
where $F: \mathbb{R}^N \times \mathbb{R}^N \to \mathbb{R}^N$. We 
introduce the shorthand $\mathbf{a} := \mathbf{x}^{t-1}$ for the input state and $\mathbf{x} := \mathbf{x}^t$ for the post-solve state. We then impose three regularity conditions. First, $F$ is twice continuously differentiable in both arguments. Second, the Jacobian $\partial_\mathbf{x} F(\mathbf{a}, \mathbf{x})$ is nonsingular at a reference pair $(\mathbf{a}, \mathbf{x})$ satisfying $F(\mathbf{a}, \mathbf{x}) = \mathbf{0}$. Finally, the analysis is restricted to a local neighborhood of this reference solution. Under these conditions, the IFT gives a locally unique and continuously differentiable map $\Phi: \mathbb{R}^N \to \mathbb{R}^N$, denoted $\mathbf{x} = \Phi(\mathbf{a})$, satisfying
\begin{equation}
F(\mathbf{a}, \Phi(\mathbf{a})) = \mathbf{0},
\label{eq:ift_identity}
\end{equation}
which induces the locally well-defined time-step operator:
\begin{equation}
\mathbf{x}^t = \Phi(\mathbf{x}^{t-1}).
\label{eq:time_step_operator}
\end{equation}
Although $\Phi$ is not available in closed form, it is evaluated 
numerically at each time step by solving the nonlinear 
system Eq.~\eqref{eq:residual_general} using the Newton--Krylov method 
described in Section~\ref{sec:timestep}.

Suppose the input state to a solver step is perturbed: $\tilde{\mathbf{a}} 
= \mathbf{a} + \eta$, where $\eta \in \mathbb{R}^N$ is an arbitrary 
perturbation. Define the unperturbed and perturbed outputs respectively as
\begin{equation*}
\mathbf{x} = \Phi(\mathbf{a}), \qquad \tilde{\mathbf{x}} = \Phi(\tilde{\mathbf{a}}).
\end{equation*}
The deviation in the output state after one time step is
\begin{equation}
e := \tilde{\mathbf{x}} - \mathbf{x} = \Phi(\mathbf{a} + \eta) - \Phi(\mathbf{a}),
\label{eq:single_step_error_def}
\end{equation}
where $e$ represents the error in the output state due to the 
perturbation $\eta$ in the input state. Since $\Phi$ is differentiable, 
a first-order Taylor expansion gives
\begin{equation}
e = S(\mathbf{a})\,\eta + r(\eta), \qquad \text{where } 
\|r(\eta)\| = \mathcal{O}(\|\eta\|^2),
\label{eq:single_step_with_remainder}
\end{equation}
where $S(\mathbf{a}) := \tfrac{d\Phi(\mathbf{a})}{d\mathbf{a}}$ is the sensitivity matrix and $r(\eta)$ is the leading-order remainder. To obtain a quantitative sensitivity bound, we take norms on both sides:
\begin{equation}
\|e\| = \|S(\mathbf{a})\,\eta + r(\eta)\|.
\label{eq:norm_both_sides}
\end{equation}
Applying the triangle inequality yields
\begin{equation}
\|e\| \leq \|S(\mathbf{a})\,\eta\| + \|r(\eta)\|.
\label{eq:triangle_inequality}
\end{equation}
Invoking the submultiplicative property on the first term and the 
remainder bound on the second gives
\begin{equation}
\|e\| \leq \|S(\mathbf{a})\|\, \|\eta\| + \mathcal{O}(\|\eta\|^2),
\label{eq:sensitivity_bound}
\end{equation}
which quantifies how the leading-order error is amplified by the 
operator norm of the sensitivity matrix $S(\mathbf{a})$.

We now derive the explicit form of $S(\mathbf{a})$. Differentiating the 
defining identity Eq.~\eqref{eq:ift_identity}, $F(\mathbf{a}, \Phi(\mathbf{a})) 
= \mathbf{0}$, with respect to $\mathbf{a}$ and applying the chain rule gives
\begin{equation}
\frac{d}{d\mathbf{a}}\bigl[F(\mathbf{a}, \Phi(\mathbf{a}))\bigr] 
= \partial_\mathbf{a} F(\mathbf{a}, \Phi(\mathbf{a})) \cdot \frac{d\mathbf{a}}{d\mathbf{a}} 
+ \partial_\mathbf{x} F(\mathbf{a}, \Phi(\mathbf{a})) \cdot \frac{d\Phi(\mathbf{a})}{d\mathbf{a}} = \mathbf{0},
\label{eq:chain_rule}
\end{equation}
where $\partial_\mathbf{a} F$ and $\partial_\mathbf{x} F$ denote the 
Fr\'echet derivatives of $F$ with respect to its first and second 
arguments, respectively. Since $\tfrac{d\mathbf{a}}{d\mathbf{a}} = I$, 
Eq.~\eqref{eq:chain_rule} simplifies to
\begin{equation}
\partial_\mathbf{a} F(\mathbf{a}, \Phi(\mathbf{a})) + \partial_\mathbf{x} F(\mathbf{a}, \Phi(\mathbf{a}))\, S(\mathbf{a}) = \mathbf{0}.
\label{eq:sensitivity_relation}
\end{equation}
Since $\partial_\mathbf{x} F(\mathbf{a}, \Phi(\mathbf{a}))$ is nonsingular 
by assumption, solving Eq.~\eqref{eq:sensitivity_relation} for $S(\mathbf{a})$ yields
\begin{equation}
S(\mathbf{a}) := -\bigl[\partial_\mathbf{x} F(\mathbf{a}, \Phi(\mathbf{a}))\bigr]^{-1}\, \partial_\mathbf{a} F(\mathbf{a}, \Phi(\mathbf{a})).
\label{eq:sensitivity_operator}
\end{equation}
Physically, $S(\mathbf{a})$ measures how strongly a single solver step
amplifies or damps a perturbation in its input. When $\|S\| > 1$ the perturbation grows across the step; when
$\|S\| < 1$ it shrinks, and the perturbed trajectory is drawn back
toward the reference trajectory. Note that $S(\mathbf{a})$ is determined entirely by the residual $F$,
which depends on the governing equations, the spatial discretization, and the time step $\tau$. It is independent of the MLT parameters $m$ and $s$ as well as the model
used to generate the prediction.
The single-step result derived here is applied recursively in
Section~\ref{sec:multi_step}, first across the steps of a single MLT
cycle and then across multiple cycles.

\subsection{Error Propagation Across Cycles}
\label{sec:multi_step}

To characterize how the error evolves over the entire simulation,
we first examine how it evolves over a single cycle.
Let $T_\ell$ denote the time index at which the ML prediction is
supplied in cycle $\ell$, which coincides with the last skipped step
(skip $s$ in Figure~\ref{fig:multi_cycle}).
The ML-initialized step (shown in blue in
Figure~\ref{fig:multi_cycle}) then runs at time index $T_\ell + 1$,
and the cycle concludes at time index $T_\ell + 1 + m$
after the $m$ physics-based correction steps.

Consider an arbitrary cycle $\ell \in \{1, \ldots, N_c\}$,
in which the ML model supplies a prediction
$\tilde{\mathbf{x}}^{T_\ell} = \mathbf{x}^{T_\ell} + \eta^{T_\ell}$
at time index $T_\ell$ in place of the true state
$\mathbf{x}^{T_\ell}$,
with $\eta^{T_\ell} \in \mathbb{R}^N$ denoting the ML prediction error.
The physics-based solver then takes $\tilde{\mathbf{x}}^{T_\ell}$
as input to the ML-initialized step at $T_\ell + 1$
and continues through the $m$ correction steps at
$T_\ell + 2, \ldots, T_\ell + 1 + m$.
Let $\{\mathbf{x}^t\}$ denote the reference trajectory satisfying
$\mathbf{x}^{t+1} = \Phi(\mathbf{x}^t)$,
and let $\{\tilde{\mathbf{x}}^t\}$ denote the perturbed trajectory
obtained by seeding the ML-initialized step with
$\tilde{\mathbf{x}}^{T_\ell}$.
The reference trajectory is taken as the state entering cycle $\ell$,
so the analysis below describes the error generated
within that cycle alone.

We define the error at step $T_\ell+j$ as
$e^{T_\ell+j} := \tilde{\mathbf{x}}^{T_\ell+j} - \mathbf{x}^{T_\ell+j}$
for $j = 1, \ldots, m+1$.
Applying the single-step result of Section~\ref{sec:single_step}
at each step of the cycle gives
\begin{equation}
e^{T_\ell+j+1} = S^{T_\ell+j+1} e^{T_\ell+j} + r^{T_\ell+j+1},
\qquad S^{T_\ell+j+1} := S(\mathbf{x}^{T_\ell+j}),
\label{eq:recursion}
\end{equation}
where $r^{T_\ell+j+1} = \mathcal{O}(\|e^{T_\ell+j}\|^2)$
is the single-step remainder obtained by Taylor-expanding $\Phi$
about the reference state $\mathbf{x}^{T_\ell+j}$
with the current error $e^{T_\ell+j}$ playing the role of the
perturbation.
The recursion is initialized by the output of the ML-initialized step,
\begin{equation}
e^{T_\ell+1} = S^{T_\ell+1} \eta^{T_\ell} + r^{T_\ell+1},
\label{eq:ml_init_step}
\end{equation}
which is the single-step result of Section~\ref{sec:single_step}
applied to the input perturbation $\eta^{T_\ell}$ at $T_\ell$.
Applying the recursion Eq.~\eqref{eq:recursion} $m$ times, starting from Eq.~\eqref{eq:ml_init_step},
yields the convolution formula
\begin{equation}
e^{T_\ell+1+m} =
\underbrace{\left(\prod_{j=1}^{m+1} S^{T_\ell+j}\right)
\eta^{T_\ell}}_{\text{initial perturbation effect }
(e^m_{\text{init}})}
+ \underbrace{\sum_{i=1}^{m+1}
\left(\prod_{j=i+1}^{m+1} S^{T_\ell+j}\right)
r^{T_\ell+i}}_{\text{accumulated remainder } (R^m)},
\label{eq:convolution}
\end{equation}
where $i$ and $j$ are summation indices local to this expression
and products with reversed limits are interpreted as the identity.
This formula decomposes the total error at the end of the cycle as
\begin{equation}
e^{T_\ell+1+m} = e^m_{\text{init}} + R^m,
\label{eq:decomposition}
\end{equation}
where $e^m_{\text{init}}$ represents the propagation of the injected ML error $\eta^{T_\ell}$ through the product of sensitivity operators along the cycle, and $R^m$ collects the single-step remainders deposited at each solver step and propagated forward by subsequent sensitivity operators. Since each error in the cycle is itself $\mathcal{O}(\|\eta^{T_\ell}\|)$, every remainder it generates is $\mathcal{O}(\|\eta^{T_\ell}\|^2)$, and
$R^m$ inherits this scaling.

Having characterized the error over a single cycle, we now turn to
the full simulation horizon, over which the ML model introduces a
fresh perturbation at the start of every cycle.
\label{sec:global_error}
Let $N$ denote the total number of time steps in the
cycle-based stage, so the simulation ends at $T_{N_c} + 1 + m = N$. Fresh
perturbations are injected at the time steps $T_1, T_2, \ldots,
T_{N_c}$, where $T_\ell$ denotes the prediction step of cycle $\ell$
as defined above, and $\eta^{T_\ell}$ is the
raw ML error injected at cycle $\ell$. Extending the single-cycle
result Eq.~\eqref{eq:decomposition} to these repeated
injections gives the modified convolution formula
\begin{equation}
e^{T_{N_c}+1+m} =
\underbrace{\left(\prod_{j=T_1+1}^{N} S^j\right)\eta^{T_1}}_{\text{initial perturbation effect } (e^N_{\text{init}})}
+
\underbrace{\sum_{\ell=2}^{N_c}\left(\prod_{j=T_\ell+1}^{N} S^j\right)\eta^{T_\ell}}_{\text{accumulated ML prediction errors } (e^N_{\text{ML}})}
+
\underbrace{\sum_{i=1}^{N}\left(\prod_{j=i+1}^{N} S^j\right)r^i}_{\text{accumulated remainder } (R^N)},
\label{eq:global_convolution}
\end{equation}
where $i$, $j$, and $\ell$ are summation indices local to this
expression. Since $T_{N_c}+1+m=N$, this formula decomposes the total
error as
\begin{equation}
e^N = e^N_{\text{init}} + e^N_{\text{ML}} + R^N,
\label{eq:global_decomposition}
\end{equation}
where $e^N_{\text{init}}$ represents the propagation of the first cycle's
injected perturbation $\eta^{T_1}$ across the full horizon,
$e^N_{\text{ML}}$ collects the contributions of all subsequent cycle
injections, and $R^N$ accumulates the single-step remainders deposited at
each solver step.
Note that unlike the single-step error, $\|S\| < 1$ alone does not
determine whether the total error decays or grows.
Each cycle injects a fresh perturbation and every solver step deposits a
new remainder, so even when $S$ damps each one individually, their
contributions can build up over many cycles.
The error at the end of the simulation therefore depends on the size of
the injected perturbations and remainders, the sensitivity $S$, and the cycle
parameters $m$ and $s$.
Section~\ref{sec:verification} examines this behavior numerically across
a range of perturbation magnitudes and cycle parameters.

\section{Numerical Verification}
\label{sec:verification}
In this section, we first confirm that the accumulated remainder
scales quadratically with the injected perturbation,
as the first-order analysis predicts.
We then apply the error propagation formula Eq.~\eqref{eq:global_convolution}
to determine when the error stays bounded across cycles,
guide the choice of correction window $m$ and skip length $s$,
and quantify the prediction accuracy
an ML model must achieve.

Throughout, the ML prediction is replaced by
a controlled synthetic perturbation of prescribed magnitude.
This isolates the error propagation behavior
from the particular ML model or data-driven method used,
so the results characterize the method itself
rather than any one predictor.
We use two testbeds, chosen so that $\|S\|$ behaves differently in each:
the pure Allen--Cahn equation Eq.~\eqref{eq:allen_cahn} and a variant with a time-dependent forcing term Eq.~\eqref{eq:forced_ac} that prevents the dynamics from settling into equilibrium. While the accuracy thresholds obtained below
are specific to these testbeds,
the procedure used to obtain them is not. $S$ follows from the solver's Jacobian and can be evaluated for any problem to assess
whether MLT is viable before an ML model is trained.

\subsection{Sensitivity Matrix Derivation for Allen--Cahn}
\label{sec:allen_cahn}

We begin by introducing the Allen--Cahn equation used
as a testbed and deriving its sensitivity matrix $S$
in closed form.
Sections~\ref{sec:single_cycle_verification} onward use this
to verify and apply the error propagation formulas
of Section~\ref{sec:theory}. The Allen--Cahn equation describes the evolution of a phase field $u$ that separates into distinct regions connected by a diffuse interface whose width is set by $\varepsilon$. On a bounded domain $\Omega \subset \mathbb{R}^d$,  it is written as
\begin{equation}
u_t = \varepsilon^2 \Delta u + u - u^3 \quad \text{in } \Omega,
\label{eq:allen_cahn}
\end{equation}
with homogeneous Neumann boundary conditions $\partial_n u = 0$ on
$\partial\Omega$, and initial condition $u(\cdot, 0) = u_0$.

We choose this equation as a testbed because it shares
the features that make time integration expensive in
many practical mechanics problems such as Navier--Stokes
and combustion with thin reaction fronts.
Each combines a diffusion term with a nonlinearity,
and a small parameter, $\varepsilon$ here or viscosity
in fluids, creates thin layers that require implicit
time integration.
Unlike more complex problems, it is a gradient flow,
which drives the pure problem toward a contractive
regime. Section~\ref{sec:forced_verification} therefore adds a forcing term to test the method outside this regime.

We discretize the Allen--Cahn equation using the implicit midpoint method. After spatial
discretization, let $\mathbf{x}(t) \in \mathbb{R}^N$ denote the discrete
state and $\Delta_h \in \mathbb{R}^{N \times N}$ the matrix discretizing the
Laplacian $\Delta$ at grid spacing $h$. The semi-discrete nonlinear operator
is
\begin{equation}
\mathcal{L}(\mathbf{z}) := \varepsilon^2 \Delta_h \mathbf{z} + \mathbf{z} - \mathbf{z}^{\circ 3},
\label{eq:semidiscrete_op}
\end{equation}
where $(\mathbf{z}^{\circ 3})_i = z_i^3$ is applied componentwise. Using the
same definitions for $F$ and $\mathbf{a}$ as in
Section~\ref{sec:single_step}, the implicit midpoint rule with time step
$\tau > 0$ gives the residual
\begin{equation}
F(\mathbf{a}, \mathbf{x}) := \mathbf{x} - \mathbf{a} - \tau \mathcal{L}\!\left(\bar{\mathbf{x}}\right) = \mathbf{0}, \qquad \bar{\mathbf{x}} = \frac{\mathbf{a}+\mathbf{x}}{2}.
\label{eq:ac_residual}
\end{equation}
The Fr\'echet derivative of $\mathcal{L}$ at $\bar{\mathbf{x}}$ is
\begin{equation}
D\mathcal{L}(\bar{\mathbf{x}}) = \varepsilon^2 \Delta_h + I - 3\,\text{diag}(\bar{\mathbf{x}}^{\circ 2}),
\label{eq:ac_frechet}
\end{equation}
and differentiating Eq.~\eqref{eq:ac_residual} with respect to $\mathbf{x}$ and
$\mathbf{a}$, using $\partial\bar{\mathbf{x}}/\partial\mathbf{x} =
\partial\bar{\mathbf{x}}/\partial\mathbf{a} = \tfrac{1}{2}I$, gives
\begin{equation}
\partial_\mathbf{x} F(\mathbf{a}, \mathbf{x}) = I - \frac{\tau}{2} D\mathcal{L}(\bar{\mathbf{x}}), \qquad \partial_\mathbf{a} F(\mathbf{a}, \mathbf{x}) = -I - \frac{\tau}{2} D\mathcal{L}(\bar{\mathbf{x}}).
\label{eq:ac_jacobians}
\end{equation}
Substituting into the general formula, Eq.~\eqref{eq:sensitivity_operator}, yields
the closed-form sensitivity matrix
\begin{equation}
S(\mathbf{a}) = \left(I - \frac{\tau}{2} D\mathcal{L}(\bar{\mathbf{x}})\right)^{-1} \left(I + \frac{\tau}{2} D\mathcal{L}(\bar{\mathbf{x}})\right), \qquad \bar{\mathbf{x}} = \frac{\mathbf{a}+\Phi(\mathbf{a})}{2}.
\label{eq:sensitivity_ac}
\end{equation}
As noted in Section~\ref{sec:single_step},
$S(\mathbf{a})$ is determined by the residual $F$ alone.
Its behavior in the following sections
therefore reflects the underlying dynamics of the problem,
not the parameters $m$ and $s$.

\subsection{Quadratic Scaling of the Cycle Remainder}
\label{sec:single_cycle_verification}

We confirm that the remainder $R^m$ in Eq.~\eqref{eq:convolution}
scales as $\mathcal{O}(\|\eta^{T_1}\|^2)$,
as predicted by the single-cycle error formula.
This is tested on the Allen--Cahn system, discretized on
$\xi \in [-\pi,\pi]$ with homogeneous Neumann boundary conditions,
$N_\xi = 512$ uniformly spaced grid points, and interface width
$\varepsilon = 0.001$. The initial condition is a randomly
generated truncated Fourier series with 128 modes normalized to $[-1,1]$, and
the time step is $\tau = 0.1$. At $T_1$, corresponding to $t = 1.5$,
we inject a perturbation $\eta^{T_1} = p\|\mathbf{x}^{T_1}\|\mathbf{v}$, where
$\mathbf{v} \in \mathbb{R}^{N_\xi}$ is a fixed unit vector obtained by
normalizing a vector with independent standard Gaussian entries, and vary
$p \in \{1\%, 2\%, 4\%, 8\%, 16\%\}$ to emulate a range of ML prediction
errors. We fix the cycle length at $c = s + 1 + m = 20$ and use the longest
available correction window, $m = 18$.
The solver runs for these $m$ correction steps after the injection, and all
quantities are evaluated at the end of this window.

\begin{figure}[!htbp]
  \centering
  \includegraphics[width=0.6\textwidth]{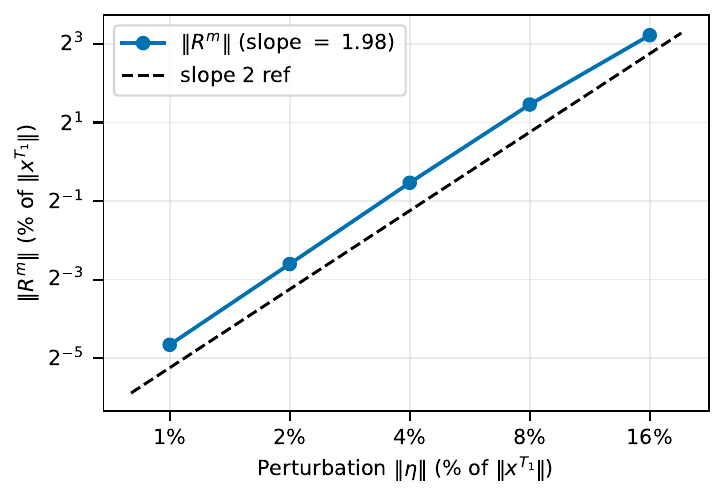}
  \caption{Quadratic scaling of $R^m$ with the injected perturbation ($m = 18$,
$\tau = 0.1$, $N_\xi = 512$); dashed line marks slope 2.}
  \label{fig:remainder_scaling}
\end{figure}

Since $R^m$ is obtained as the difference between the measured error
and the first-order prediction $e^m_{\text{init}}$,
an error in the sensitivity matrix would leave a first-order component
in this difference and pull the measured slope toward unity.
Figure~\ref{fig:remainder_scaling} instead gives a fitted slope of $1.98$
against the theoretical value of $2$,
confirming both the derivation of $S$ and its implementation.
We restrict this test to $p \le 16\%$, since quadratic scaling is expected only for small perturbations.
At larger perturbations, higher-order terms may become significant and the remainder may depart from the slope-2 trend.
The multi-cycle formula, Eq.~\eqref{eq:global_convolution}, is built
from the same single-step expansion, so this verification underpins
its use over the full simulation horizon in the sections that follow.

\subsection{Error Propagation in the Pure Allen--Cahn Equation}
\label{sec:global_verification}

We now use the multi-cycle formula, Eq.~\eqref{eq:global_convolution},
to study how error propagates over the full simulation horizon.
Fresh ML perturbations are injected at every prediction
step $T_\ell$. All experiments use $\tau = 0.1$ and $T_{\text{final}} = 20.0$.
The warm-up phase runs for $W = 15$ steps ($t = 1.5$).
This is long enough to develop a physically consistent trajectory
from the random initial condition, yet short enough that cycles
begin while the dynamics are still active, in the sense that a
perturbation is amplified rather than damped over a step
($\|S\| > 1$; see Section~\ref{sec:forced_verification}). The cycle length is fixed at $c = s + 1 + m = 20$, giving $9$ complete cycles and room for a wide range of $(m, s)$ splits. We test $(m, s) \in \{(1,18), (3,16), (5,14), (7,12), (10,9), (12,7), (14,5), (16,3), (18,1)\}$ and $p \in \{1\%, 4\%, 16\%, 32\%\}$, with perturbation $\eta^{T_\ell} = p\|\mathbf{x}^{T_\ell}\|\mathbf{v}$
constructed as in Section~\ref{sec:single_cycle_verification}.

\begin{figure}[!htbp]
  \centering

  \begin{subfigure}[b]{0.48\textwidth}
    \centering
    \includegraphics[width=\linewidth]{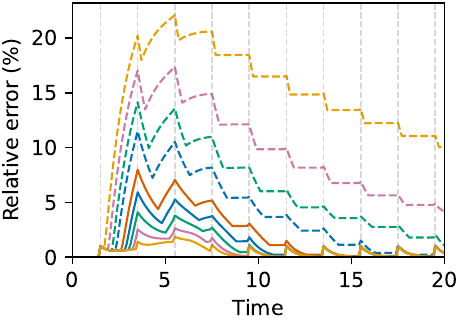}
    \caption{$\|\eta^{T_\ell}\| = 1\%$ of reference state}
    \label{fig:ms_pure_a}
  \end{subfigure}
  \hfill
  \begin{subfigure}[b]{0.48\textwidth}
    \centering
    \includegraphics[width=\linewidth]{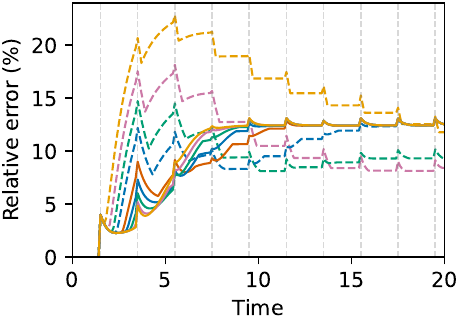}
    \caption{$\|\eta^{T_\ell}\| = 4\%$ of reference state}
    \label{fig:ms_pure_b}
  \end{subfigure}
  \vspace{0.2em}
  \begin{subfigure}[b]{0.48\textwidth}
    \centering
    \includegraphics[width=\linewidth]{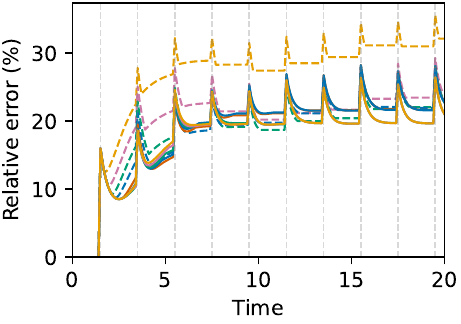}
    \caption{$\|\eta^{T_\ell}\| = 16\%$ of reference state}
    \label{fig:ms_pure_c}
  \end{subfigure}
  \hfill
  \begin{subfigure}[b]{0.48\textwidth}
    \centering
    \includegraphics[width=\linewidth]{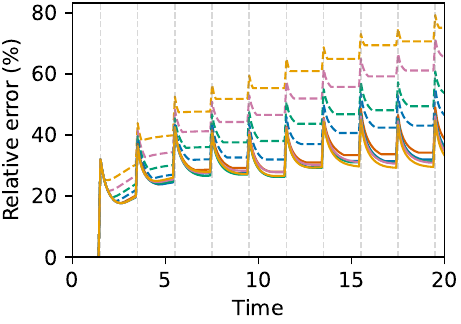}
    \caption{$\|\eta^{T_\ell}\| = 32\%$ of reference state}
    \label{fig:ms_pure_d}
  \end{subfigure}
  \vspace{0.2em}
  \includegraphics[width=\textwidth]{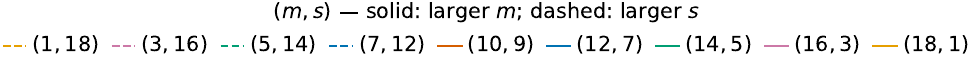}

  \caption{Effect of $(m,s)$ on error propagation for pure Allen--Cahn,
with $p$ increasing from (a) to (d), where the injected perturbation is
$\eta^{T_\ell} = p\|\mathbf{x}^{T_\ell}\|\mathbf{v}$.
Solid lines: larger $m$; dashed lines: larger $s$.}
  \label{fig:ms_effect_pure}
\end{figure}

Figure~\ref{fig:ms_effect_pure} shows the relative error over the
simulation for each $(m,s)$ combination, at four values of $p$ increasing
from Figures~\ref{fig:ms_pure_a}--\ref{fig:ms_pure_d}.
Each cycle produces a spike at the ML-initialized step, before the
correction steps have acted.
The quantity of interest is the lower envelope traced once the corrections have taken effect, and statements below about whether the error decays, plateaus, or grows refer to that envelope.

With $S$ fixed by the problem setup, whether the error remains bounded depends on $p$ and the correction window.
At $p = 1\%$ (Figure~\ref{fig:ms_pure_a}) the error decays once the dynamics contract, and at $p = 4\%$ (Figure~\ref{fig:ms_pure_b}) it flattens. At $p = 16\%$ (Figure~\ref{fig:ms_pure_c}) it plateaus for all but two of the most skip-heavy combinations, $(m=1,s=18)$ and $(m=3,s=16)$, where it drifts slowly upward. At $p = 32\%$ (Figure~\ref{fig:ms_pure_d}), configurations with $m \leq 5$ grow steadily through the end of the simulation. The rest remain bounded. In summary, prediction errors up to $p = 16\%$ are absorbed provided $m \geq 5$, and at $p = 32\%$ provided $m \geq 7$. At $p = 32\%$, however, the bounded error settles at $30\%$ to $40\%$, which may exceed what an application can tolerate. The accuracy an ML model must achieve for this problem therefore depends on these bounds and on the error the application can tolerate.

\begin{figure}[!htbp]
\centering
\makebox[\textwidth][c]{\includegraphics[width=1.15\textwidth]
{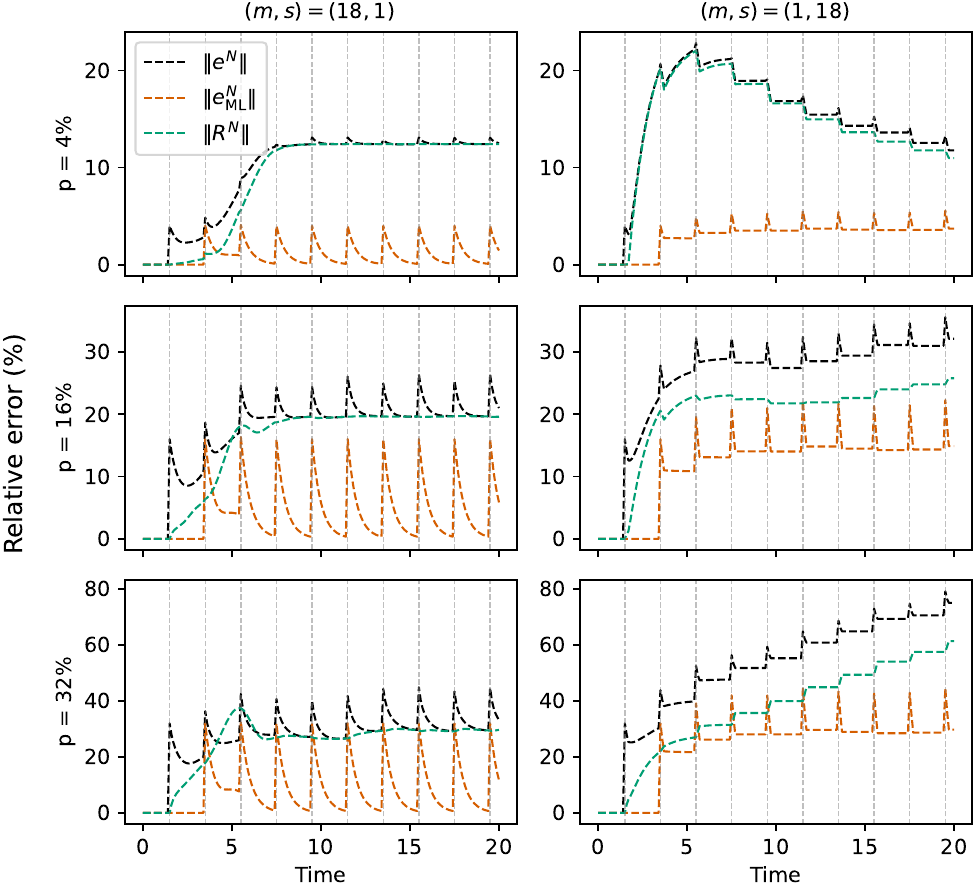}}
\caption{Error decomposition for $(m,s) = (18,1)$ and $(1,18)$ at $p \in \{4\%, 16\%, 32\%\}$, where the injected perturbation is $\eta^{T_\ell} = p\|\mathbf{x}^{T_\ell}\|\mathbf{v}$. Vertical lines mark ML injection points.}
\label{fig:ms_decomp}
\end{figure}

At $p = 4\%$ (Figure~\ref{fig:ms_pure_b}) an anomaly appears after $t \approx 5$. Skip-heavy combinations perform worse initially but decline stepwise
until they overtake some larger-$m$ combinations.
The cause is visible in the two error terms that remain
in Eq.~\eqref{eq:global_convolution} once the initial perturbation effect has decayed.
Figure~\ref{fig:ms_decomp} separates them for the extreme combinations $(m,s) = (18,1)$ and $(1,18)$.
At $p = 4\%$ (top row), the 18 correction steps in a cycle of $(m=18,s=1)$ remove $e^N_{\text{ML}}$.
These steps also keep depositing remainders, so $R^N$ sets the level of the error. With $(m=1,s=18)$, a single correction step cannot remove $e^N_{\text{ML}}$, but far fewer remainders are deposited. $R^N$ therefore declines stepwise, bringing the total error down to the
level of the larger-$m$ combinations by the end of the simulation. This accounts for the overtaking seen in Figure~\ref{fig:ms_pure_b}. At $p = 16\%$ (middle row) and $p = 32\%$ (bottom row) the unremoved $e^N_{\text{ML}}$ of $(1,18)$ is larger than at $p = 4\%$, which inflates the remainder at every solver step. Contraction by $S$ then can no longer bring $R^N$ down, and at $p = 32\%$ $R^N$ grows, reversing the ordering.

In summary, the physics-based correction therefore keeps the error bounded whenever the window is long enough, even at the largest perturbations tested. Larger correction windows generally give lower error, though the gain diminishes at the upper end. For example, $(m=16,s=3)$ and $(m=18,s=1)$ produce nearly identical results.
In practice $m$ should be chosen just large enough to suppress the injected error within a cycle, since increasing it further costs speedup without much gain in accuracy. The thresholds themselves are specific to Allen--Cahn since $S$ is fixed by the problem setup. However, the method used for the analysis is general and can be applied to any problem whose solver Jacobian is available, giving the corresponding thresholds before an ML model is trained.
Section~\ref{sec:forced_verification} tests whether the guidance provided here holds under more dynamically active conditions.

\captionsetup[subfigure]{skip=2pt}

\subsection{Error Propagation with Dynamic Forcing}
\label{sec:forced_verification}

The experiments of Section~\ref{sec:global_verification} used the pure Allen--Cahn equation, which approaches a steady solution once the phases separate. To assess the robustness of the guidance derived there, we repeat the analysis on a testbed where the dynamics persist, the forced Allen--Cahn equation
\begin{equation}
u_t = \varepsilon^2 \Delta u + u - u^3 + f(\xi, t),
\label{eq:forced_ac}
\end{equation}
where the forcing term
\begin{equation}
f(\xi, t) = \sum_{i=1}^{10} A_i \sin(k_i \xi - \omega_i t), \quad
k \in \{2, 3, 4, 6, 10, 12, 15, 18, 20, 25\},
\label{eq:forcing}
\end{equation}
is a superposition of ten traveling waves with wavenumbers from 2 to 25, amplitudes $A_i \in [0.08, 0.20]$ and frequencies $\omega_i \in [0.5\pi, 3\pi]$, chosen to keep the dynamics from settling. All other parameters are identical to Section~\ref{sec:global_verification}.

\begin{figure}[!htbp]
\centering
\includegraphics[width=0.8\textwidth]
{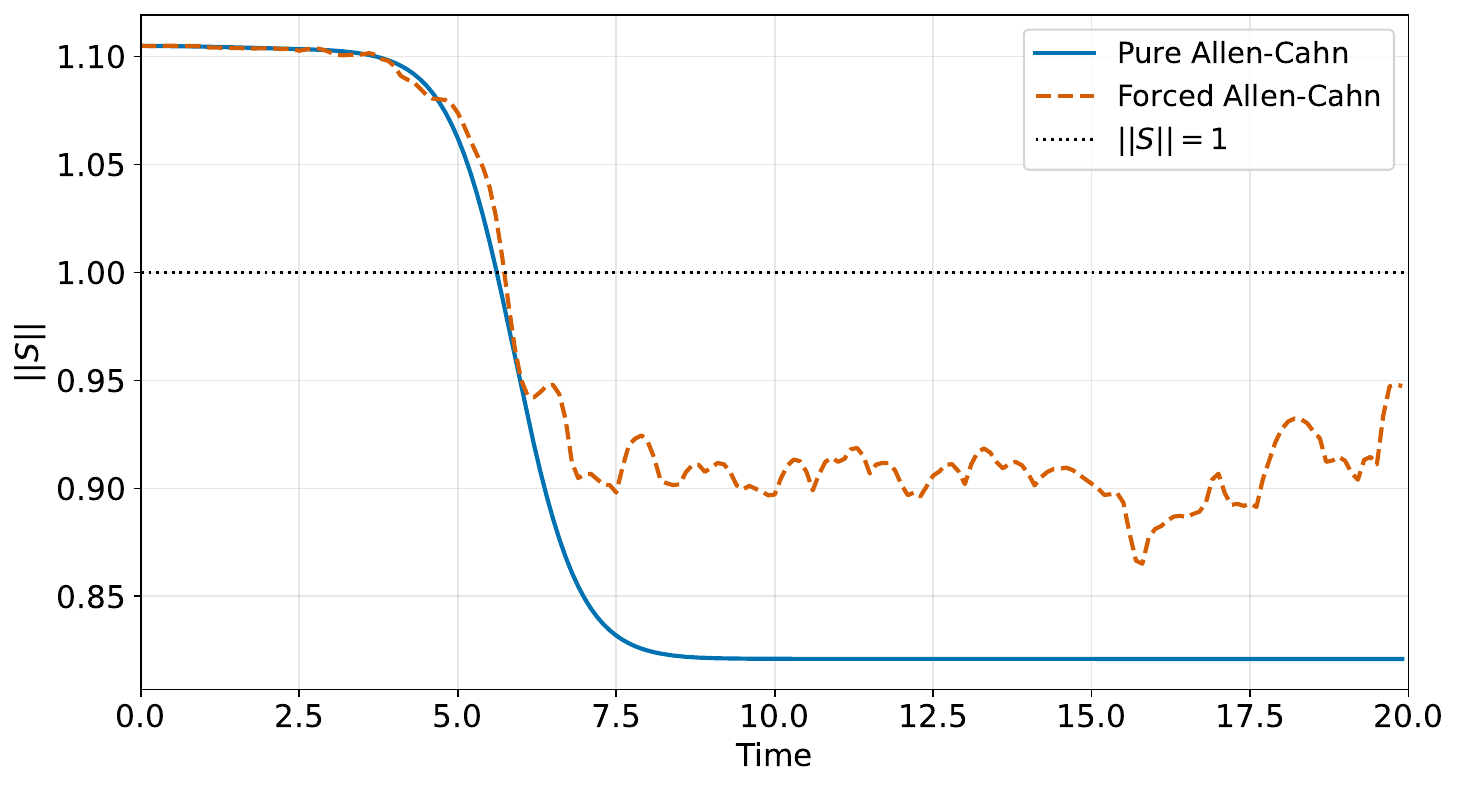}
\caption{Sensitivity operator norm $\|S\|$ for pure Allen--Cahn 
(blue solid) and forced Allen--Cahn (orange dashed) with $\tau = 
0.1$, $\varepsilon = 0.001$, $N_\xi = 512$; dotted line marks 
$\|S\| = 1$.}
\label{fig:sensitivity_comparison}
\end{figure}

Figure~\ref{fig:sensitivity_comparison} shows $\|S\|$ over the course of the simulation for both testbeds.
Both begin above unity and fall below it shortly after $t \approx 5$. However, in the pure case, $\|S\|$ then settles near $0.82$ once the phases separate, while in the forced case it never settles, oscillating between $0.87$ and $0.95$ for the remainder of the simulation. Because $S$ characterizes the problem itself, the two testbeds shape error propagation differently even for the same $m$, $s$ and
perturbation size.

\begin{figure}[!htbp]
  \centering

  \begin{subfigure}[b]{0.48\textwidth}
    \centering
    \includegraphics[width=\linewidth]{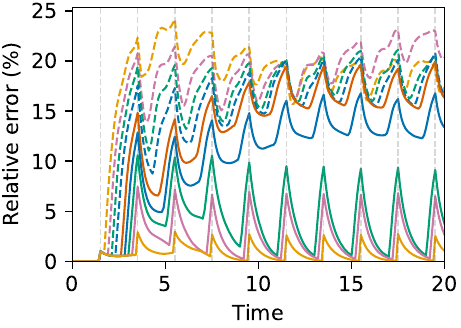}
    \caption{$\|\eta^{T_\ell}\| = 1\%$ of reference state}
    \label{fig:ms_forced_a}
  \end{subfigure}
  \hfill
  \begin{subfigure}[b]{0.48\textwidth}
    \centering
    \includegraphics[width=\linewidth]{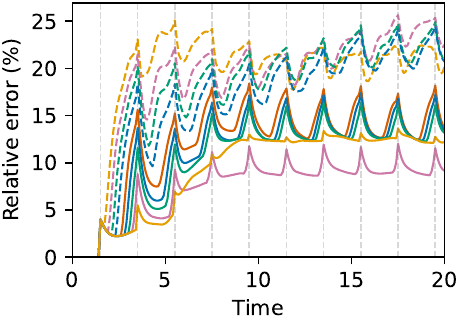}
    \caption{$\|\eta^{T_\ell}\| = 4\%$ of reference state}
    \label{fig:ms_forced_b}
  \end{subfigure}
  \vspace{0.2em}
  \begin{subfigure}[b]{0.48\textwidth}
    \centering
    \includegraphics[width=\linewidth]{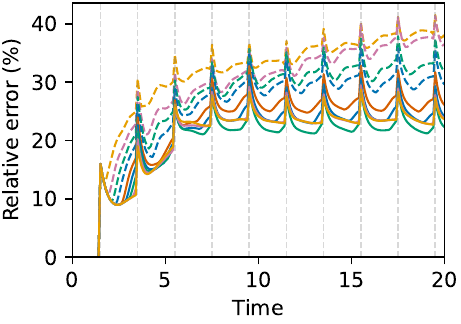}
    \caption{$\|\eta^{T_\ell}\| = 16\%$ of reference state}
    \label{fig:ms_forced_c}
  \end{subfigure}
  \hfill
  \begin{subfigure}[b]{0.48\textwidth}
    \centering
    \includegraphics[width=\linewidth]{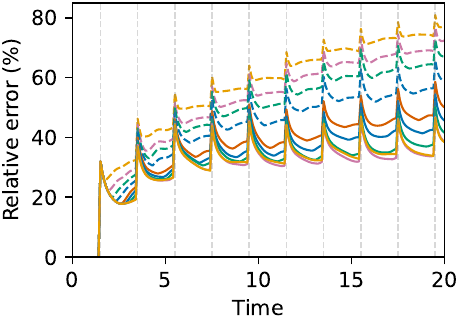}
    \caption{$\|\eta^{T_\ell}\| = 32\%$ of reference state}
    \label{fig:ms_forced_d}
  \end{subfigure}
  \vspace{0.2em}
  \includegraphics[width=\textwidth]{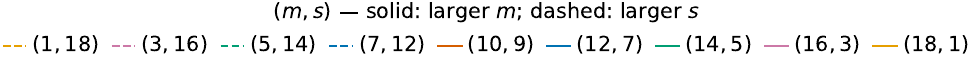}

  \caption{Effect of $(m,s)$ on error propagation for forced
Allen--Cahn, with $p$ increasing from (a) to (d). The injected
perturbation is $\eta^{T_\ell} = p\|\mathbf{x}^{T_\ell}\|\mathbf{v}$.
Solid lines: larger $m$; dashed lines: larger $s$.}
  \label{fig:ms_effect_forced}
\end{figure}

Figure~\ref{fig:ms_effect_forced} repeats the sweep of
Figure~\ref{fig:ms_effect_pure} for the forced problem.
At $p = 1\%$, $4\%$ and $16\%$ (Figures~\ref{fig:ms_forced_a}--\ref{fig:ms_forced_c}) every
configuration with $m \geq 5$ settles to a plateau, at a level that rises with $p$. At $p = 32\%$ (Figure~\ref{fig:ms_forced_d}) the larger windows still
plateau, while those with $m \leq 7$ climb through the end of the simulation. In Figures~\ref{fig:ms_forced_a}--\ref{fig:ms_forced_d}, correction steps beyond a certain window length no longer lower the error. For example, $(16,3)$ and $(18,1)$ settle at almost the same level.
The guidance of Section~\ref{sec:global_verification} therefore survives, including the recommendation to set $m$ just large enough rather than as large as possible.
We next quantify the computational speedup of the MLT method.

\subsection{Wall-Clock Speedup}
\label{sec:wall_speedup}
In this subsection, we evaluate the computational savings of the method.
We exclude the warm-up phase from this analysis.
In each cycle of length $c = s+1+m$, the solver is invoked on $m+1$ steps:
one ML-initialized step and $m$ correction steps.
The remaining $s$ steps are skipped entirely.
The theoretical speedup factor is therefore
\begin{equation}
\text{Speedup} = \frac{c}{m+1} = \frac{c}{c-s}.
\label{eq:speedup}
\end{equation}

However, when performing MLT, the introduction of the ML prediction may require extra iterations at the ML-initialized step and subsequently the correction steps.  This, therefore, reduces the actual savings compared to the theoretical value. To assess how much of the skipping translates to wall-clock performance, we measured execution time for each $(m, s)$ configuration on both the pure and forced Allen--Cahn problems, using the minimum across 10 runs to mitigate
system noise. All perturbation vectors were precomputed, so that only solver computation is timed. Timing is also restricted to the 9 complete cycles, so that the theoretical formula~\eqref{eq:speedup} applies exactly. We use $p = 4\%$ throughout, representative of a viable operating point for both problems.

\begin{figure}[!htbp]
\centering
\includegraphics[width=0.8\textwidth]{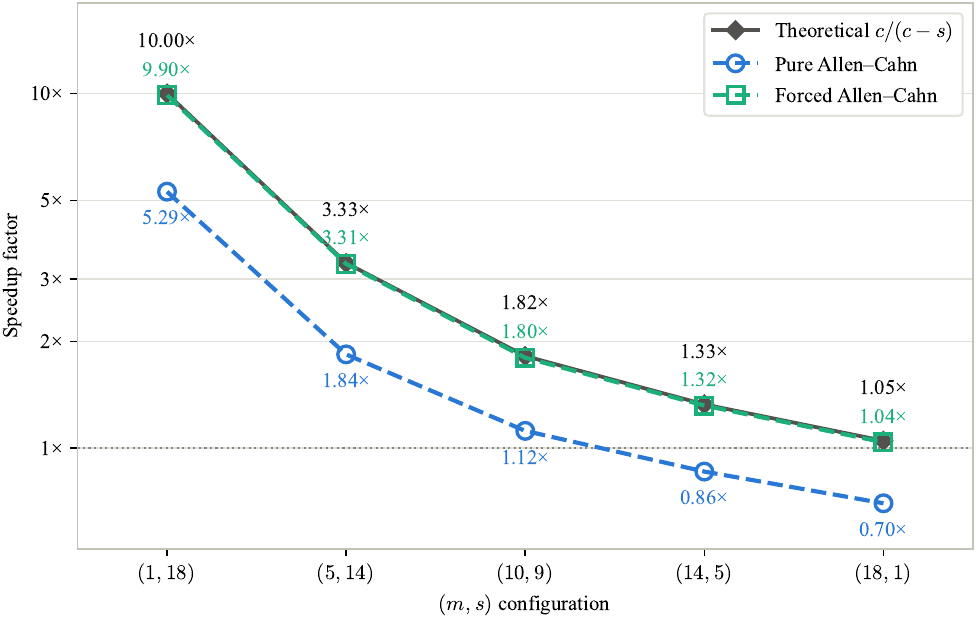}
\caption{Wall-clock speedup versus $(m,s)$ for the pure and forced
Allen--Cahn problems, with the theoretical speedup $c/(c-s)$ of
Eq.~\eqref{eq:speedup} for comparison.}

\label{fig:wall_speedup}
\end{figure}

Figure~\ref{fig:wall_speedup} compares the measured and theoretical
speedups.
For the pure case, the measured speedup falls systematically below the
theoretical value, dropping below break-even at $(14,5)$ and $(18,1)$,
where it reaches $0.86\times$ and $0.70\times$.
The cause is that the ML-initialized step is seeded with a perturbed
state, so Newton requires more iterations to converge there than at a
baseline step.

A Newton solve for pure Allen--Cahn is cheap ($\approx 6.7$\,ms), so the
extra iterations consume a large share of what skipping saves.
Configurations that skip aggressively still gain, with $(1,18)$ reaching
$5.29\times$, but those with few skipped steps save too little to offset
the penalty.
For the forced problem~\eqref{eq:forced_ac}, the Newton solve is roughly
twice as expensive ($\approx 12.8$\,ms), so the same extra iterations cost
proportionally less and the measured speedup tracks the theoretical value
across all configurations.
As the solver becomes more expensive, the penalty shrinks relative to the
work saved.
We therefore regard the theoretical speedup~\eqref{eq:speedup} as an
appropriate characterization of MLT's acceleration potential for problems
with expensive solvers.

Together with the accuracy results, this completes the trade-off
governing the choice of $m$.
The correction window must be long enough to suppress the injected
error, but every additional correction step reduces the speedup, now
quantified by Figure~\ref{fig:wall_speedup}.
Since the accuracy benefit saturates while the cost does not, $m$ should
be set just large enough to keep the error bounded.

\newpage
\section{Conclusions}
\label{sec:conclusion}

In this work, we presented a cycle-based Machine Learning-enabled Time-skipping
(MLT) method. The method reduces time-stepping cost by skipping time steps and
bridging the resulting gaps with an ML model, while the physics-based solver
corrects the resulting error. Using the Implicit Function Theorem, we derived first-order error propagation estimates, which identify the main factors controlling the
error. These are the size of the injected perturbation, the solver's
sensitivity to it, and the length of the correction window. 
Numerical verification on the Allen--Cahn equation and Allen--Cahn equation with dynamic forcing using controlled synthetic perturbations to emulate ML errors confirmed these
estimates. In both cases, the error stays bounded whenever the correction window is long enough.
Wall-clock measurements show that the theoretical time saving is
achievable when the solver is expensive enough to absorb the added cost
of the ML-initialized step. The thresholds obtained are specific to the problems studied here, but the
procedure is not.
The sensitivity follows from the solver's Jacobian and can be evaluated
for any problem.
It therefore gives the required correction window before an ML model is
trained.
Our next steps include integrating trained ML models into the time-skipping framework and extending the method to problems such as the Navier--Stokes
equations.
Furthermore, setup parameters such as cycle lengths and the number of skipped time steps can be optimized based on the nature of the problem.

\section*{CRediT authorship contribution statement}

\textbf{Isaac C. Bannerman:} Conceptualization, Methodology, Software,
Formal analysis, Investigation, Validation, Visualization,
Writing -- original draft.
\textbf{Shaowu Pan:} Conceptualization, Methodology, Supervision,
Writing -- review \& editing.
\textbf{Lucy T. Zhang:} Conceptualization, Methodology, Supervision,
Funding acquisition, Project administration, Resources,
Writing -- review \& editing.

\section*{Declaration of competing interest}

The authors declare that they have no known competing financial interests or
personal relationships that could have appeared to influence the work reported
in this paper.

\section*{Data availability}

The code and data supporting the findings of this study are available from the
corresponding author upon reasonable request.

\section*{Acknowledgments}

This work was partially supported by the National Science Foundation under
Grant No. 2033977. Any opinions, findings, and conclusions or recommendations
expressed in this material are those of the authors and do not necessarily
reflect the views of the National Science Foundation.

\section*{Declaration of generative AI and AI-assisted technologies in the
manuscript preparation process}

During the preparation of this work the authors used Claude (Anthropic) in
order to improve the clarity and readability of the text. After using this
tool, the authors reviewed and edited the content as needed and take full
responsibility for the content of the published article.
\clearpage
\bibliographystyle{unsrt}
\bibliography{reference}

\end{document}